\documentclass[aps,onecolumn,showpacs,superscriptaddress,notitlepage]{revtex4-1}
\usepackage{makeidx}
\usepackage[utf8]{inputenc}
\usepackage{graphicx}
\usepackage{amsmath}
\usepackage{mathrsfs}
\usepackage{graphicx}
\usepackage{braket}
\usepackage[subrefformat=parens,labelformat=parens,caption=false]{subfig}
\usepackage{dcolumn}
\usepackage{bm}
\usepackage{bbm}
\usepackage{color}
\usepackage[dvipsnames]{xcolor}
\usepackage{esint}
\usepackage{lineno}
\usepackage{adjustbox}
\usepackage{verbatim}
\usepackage[breaklinks,
            colorlinks,
            urlcolor=blue,
            linkcolor=blue,
            anchorcolor=blue,
            citecolor=blue]{hyperref}

\usepackage{soul}

\begin{document}

\title{Microtoroidal resonators enhance long-distance dynamical entanglement generation in chiral quantum networks}

\author{Wai-Keong Mok}
\affiliation{Centre for Quantum Technologies, National University of Singapore, 3 Science Drive 2, Singapore 117543}
\email{waikeong\_mok@u.nus.edu}
\author{Davit Aghamalyan}
\affiliation{Centre for Quantum Technologies, National University of Singapore, 3 Science Drive 2, Singapore 117543}
\author{Jia-Bin You}
\affiliation{Department of Electronics and Photonics, Institute of High Performance Computing, 1 Fusionopolis Way, 16-16 Connexis,
Singapore 138632, Singapore}
\author{Leong-Chuan Kwek}
\affiliation{Centre for Quantum Technologies, National University of Singapore, 3 Science Drive 2, Singapore 117543}
\affiliation{MajuLab, CNRS-UNS-NUS-NTU International Joint Research Unit, UMI 3654, Singapore}
\affiliation{National Institute of Education and Institute of Advanced Studies,
Nanyang Technological University, 1 Nanyang Walk, Singapore 637616}

\begin{abstract}

Chiral quantum networks provide a promising  route for realising quantum information processing and quantum communication. Here, we describe how two distant quantum nodes of chiral quantum network  become dynamically entangled by a photon transfer through a common 1D chiral waveguide. We harness the directional asymmetry in chirally-coupled single-mode ring resonators to generate  entangled state between two atoms. We report a concurrence of up to 0.969, a huge improvement over the 0.736 which was suggested and analyzed in great detail in Ref. \cite{gonzalez2015chiral}. This significant enhancement is achieved by introducing microtoroidal resonators which serve as efficient photonic interface between light and matter.  Robustness of our protocol to experimental imperfections such as fluctuations in inter-nodal distance, imperfect chirality, various detunings and atomic spontaneous decay is demonstrated.  Our proposal can be utilised for long-distance entanglement generation in quantum networks which is a key ingredient for many applications in quantum computing and quantum information processing.

\end{abstract}

\maketitle

\section{Introduction}
%Quantum networks
Quantum networks \cite{Kimble2008,  RR2015} provide an elegant solution for addressing crucial tasks of quantum computing and quantum communication. Quantum network typically comprises nodes, which are linked together via the quantum bus, utilizing `flying qubits' such as photons which are fast information carriers. The interaction between qubit and cavity paves the way to realize quantum gates which are elementary building blocks for quantum information processing and computing \cite{Nielsen2000}. Quantum networks are thus promising for the future implementation of quantum computation, communication, and metrology \cite{Nielsen2000,Zoller2005,Gisin2007,Giovannetti1330,Giovanetti2010}.

In the optical regime, cold atoms in Fabry-Perot cavities connected via waveguides have been demonstrated to be good candidates for a quantum network~\cite{Raimond2001,BPK12,cirac1997quantum,RR2015,WV2006,ritter2012elementary}. However, these implementations suffer from the drawback of scalability. To overcome this problem, several types of microchip-based systems (microdisk, micropillar, micro bottle, and photonic crystal cavities) \cite{Vahala2004} have been engineered and successfully utilized for creating light-matter interfaces, which are used in cavity QED experiments by coupling them with trapped cold atoms and quantum dots \cite{A2006,S2006,Spillane2005,A2009,S2007,OJV2013,kippenberg2005,S2014,Dayan2008}. Microtoroidal and microdisk cavities have the huge potential to realize scalable quantum networks. These so-called whispering gallery mode (WGM) resonators are typically dielectric spherical structures in which light is confined due to the total internal reflection. As a consequence of their small losses, these systems have high quality factors as large as $4\times 10^{8}$ \cite{kippenberg2005}. Moreover, these resonators have been experimentally demonstrated to couple to tapered fibres with a high efficiency of 0.997 \cite{A2006}. In this regard, WGM resonators are perfect candidates for realising light-matter interface.

In the context of quantum networks, an outstanding challenge is the long-distance high-fidelity entanglement generation and transfer despite having noise and dissipation present in the quantum channel \cite{Nielsen2000}.
As has been demonstrated in the seminal work by Cirac \textit{et al.} \cite{cirac1997quantum}, long-distance high-fidelity state transfer in quantum network requires the nodes to be coupled in a unidirectional fashion. Such systems are known in quantum optics as cascaded systems. Interestingly, chiral systems realize the most natural implementation of cascaded systems \cite{PhysRevLett.70.2273,gardiner1993driving,carmichael2007statistical}, where two subsystems are coupled unidirectionally without information backflow. These systems, even when they are separated by long distances, can be described under the Born-Markov approximation as the retardation effects can be accounted for by a simple redefinition of the time and phase of the target node. \cite{gardiner1993driving,carmichael2007statistical}. In recent years, chiral quantum networks have been extensively studied \cite{lodahl2017chiral,ramos2016non,PhysRevA.91.042116,ramos2016non,mahmoodian2016quantum,buonaiuto2019dynamical} and shown to be fruitful for generating pure multipartite entangled steady states \cite{PhysRevA.91.042116}, implementing universal quantum computation with heralded two-qubit gates \cite{mahmoodian2016quantum}, studying new exotic many-body phases such as the formation of chiral dimers \cite{ramos2016non,buonaiuto2019dynamical} and in the transport of maximally entangled states \cite{mok2019long}. Moreover, chiral waveguides have been shown to provide a particularly suitable platform for generating entanglement \cite{gonzalez2015chiral}. It was argued that the maximum achievable concurrence between two atoms is 1.5 times higher as compared to the non-chiral systems \cite{zheng2013persistent,mirza2016two,gonzalez2014generation,facchi2016bound,martin2011dissipation,gonzalez2011entanglement}, with the added feature that the generated entanglement is insensitive to the distance between the atoms.

%Chiral systems review
In atom-waveguide coupled systems, the chiral light-matter interaction emerges when the symmetry of photon emission in the left and right directions is broken \cite{lodahl2017chiral}. For instance, in subwavelength-diameter optical fiber, strong transverse confinement imposes selection rules between the local polarization and propagation direction of the photons. As a result, the emission and absorption of photons depend on the propagation direction. Many experiments have reported chiral systems with very large directionality \cite{young2015polarization,le2015nanophotonic,sollner2015deterministic}. In particular, photonic crystals turn out to be particularly promising as directionality of $90\%$ have been obtained in these systems \cite{young2015polarization,le2015nanophotonic}. Moreover, in photonic waveguides, the cavity decay into non-waveguide modes can be neglected by exploiting photonic bandgap effects \cite{arcari2014near}, leading to a $\beta$-factor (the ratio of emission rate into the waveguide modes to the total emission rate) close to 1.

Motivated by Ref. \cite{gonzalez2015chiral}, in this paper we study the entanglement generation between two qubits in a chiral quantum network separated by long distances. By obtaining both analytical and numerical solutions, we look for an optimal parameter regime which maximizes the two-qubit concurrence \cite{Zurek}. Each node of our chiral quantum network consists of a ring cavity externally coupled to an atom. We show that by introducing ring cavities, the concurrence between the atoms can be enhanced by additional factor of $20\%$ compared to the previous proposal outlined in Ref.~\cite{gonzalez2015chiral}. The resonators provide an additional level of control over the system dynamics, which aid significantly in providing the necessary conditions for strong entanglement generation, giving a maximum concurrence of up to $0.969$.

Compared to other schemes, our minimal control proposal has various advantages. Firstly, the scheme works in the weak coupling regime with no time-dependent control required, allowing our proposal to be easily implemented in current photonic systems. Also, the optimal entanglement generation occurs dynamically, which makes it a much faster protocol compared to steady state schemes \cite{PhysRevA.91.042116}. Additionally, the entanglement generation is insensitive to the distance between the nodes due to the cascaded nature of the network, and is robust against realistic experimental imperfections. These advantages highlight the suitability of our proposal for long-distance entanglement generation.

The paper is outlined as follows. In Sec. \ref{sec:model}, we introduce the model of a chiral quantum network which is studied analytically in Sec. \ref{sec:cavity_improve_entanglement}, where we highlight our main result that adding resonators enhance entanglement generation. Sec. \ref{sec:imperfections} looks at the robustness of our protocol against various experimental imperfections. Further optimization results are presented in Sec. \ref{sec:chiral_unequal} which gives a maximum concurrence of $0.969$. We consider an alternative non-chiral protocol in Sec. \ref{sec:non_chiral} using the same setup, which emphasises the key advantages of using a chiral waveguide. Finally, we conclude our results in Sec. \ref{sec:conclusion}.
 
\section{Model of chiral quantum network}
\label{sec:model}
The quantum network model comprises two nodes on a waveguide characterized by the wave number $k$. Each node comprises one qubit coupled to a single-mode cavity. Throughout this paper, we will refer to the left node as node 1 (at position $x_1$) and the right node as node 2 (at position $x_2$). The nodes are separated by a distance $D \equiv |x_2 - x_1|$. The master equation is given by \cite{PhysRevA.91.042116,mok2019long}:
	\begin{equation}
	\begin{split}
\dot{\rho} &= -i[H_{\text{JC}}, \rho] + \sum_{i=L,R} \sum_{j=1,2} \gamma_{ij} \mathcal{D}[a_j]\rho + \sum_{j=1,2} \Gamma_j \mathcal{D}[\sigma_j]\rho \\
&+ \sqrt{\gamma_{R1}\gamma_{R2}} (e^{-ikD} [a_2, \rho a_1^\dag] + e^{ikD} [a_1 \rho, a_2^\dag]) + \sqrt{ \gamma_{L1}\gamma_{L2}} (e^{-ikD} [a_1,\rho a_2^\dag] + e^{ikD} [a_2 \rho, a_1^\dag])
	\end{split}
	\label{eq:model}
	\end{equation}
where 
	\begin{equation}
H_{\text{JC}} = \omega_{c1} a_1^\dag a_1 + \omega_{c2} a_2^\dag a_2 + \omega_{a1} \sigma_1^+ \sigma_1 + \omega_{a2} \sigma_2^\dag \sigma_2 + g_1 (a_1^\dag \sigma_1 + a_1 \sigma_1^+) + g_2 (e^{i \alpha} a_2^\dag \sigma_2 + e^{-i \alpha} a_2 \sigma_2^+)
	\end{equation}
is the usual Jaynes-Cummings Hamiltonian describing two separate nodes, and $\mathcal{D}[\mathcal{O}]\rho = \mathcal{O}\rho \mathcal{O}^\dag - \frac{1}{2} \{ \mathcal{O}^\dag \mathcal{O}, \rho \}$ is the standard Lindblad superoperator. The annihilation operator of the cavity in node $j$ is given by $a_j$, with the corresponding atomic lowering operator $\sigma_j$. The decay rates of the cavity in node $j$ (into the waveguide) are given by $\gamma_{ij}$, where $i = L,R$ denote the emission direction. The atomic spontaneous decay rates (into non-waveguide modes) are given by $\Gamma_j$. The cavity frequencies in each node are given by $\omega_{c1}$ and $\omega_{c2}$ respectively, with the atomic frequencies $\omega_{a1}$ and $\omega_{a2}$. Cavity-atom interaction strength in node $j$ is given by $g_j$, assumed to be real, with a relative phase $\alpha$ between the two cavity-atom couplings. We note that it is straightforward to recover the simplified model considered in Ref. \cite{gonzalez2015chiral} without cavities, by dropping the terms with $\sigma_j$ in Eq. (\ref{eq:model}) and replacing $a_j \to \sigma_j$ for $j = 1,2$. 

\section{Improved entanglement generation due to cavities}
\label{sec:cavity_improve_entanglement}

We first consider the case where both cavities are symmetrically coupled to the waveguide, i.e. $\gamma_{R1} = \gamma_{R2} = \gamma_R$ and $\gamma_{L1} = \gamma_{L2} = \gamma_L$. The master equation simplifies to give
	\begin{equation}
	\begin{split}
\dot{\rho} &= -i \bigg[H_{\text{JC}} -i\frac{\gamma_L}{2} (e^{ikD} a_1^\dag a_2 - e^{-ikD} a_2^\dag a_1) -i \frac{\gamma_R}{2} (e^{ikD} a_2^\dag a_1 - e^{-ikD} a_1^\dag a_2), \rho\bigg] \\
&+ \gamma_L \mathcal{D}[e^{ikx_1} a_1 + e^{ikx_2} a_2] \rho + \gamma_R \mathcal{D}[ e^{-ikx_1} a_1 + e^{-ikx_2} a_2]\rho + \Gamma_1 \mathcal{D}[\sigma_1] \rho + \Gamma_2 \mathcal{D}[\sigma_2] 
	\label{eq:lindblad_ME}
	\end{split}
	\end{equation}
Assuming the system dynamics is restricted to a single-excitation subspace, the jump term in the master equation can be neglected to obtain the non-Hermitian Hamiltonian:
	\begin{equation}
H_{\text{eff}} = H_{\text{JC}} -i \gamma_L e^{ikD} a_1^\dag a_2 - i\gamma_R e^{ikD} a_2^\dag a_1 - i \frac{\gamma_L + \gamma_R}{2} (a_1^\dag a_1 + a_2^\dag a_2) - i \frac{\Gamma_1}{2} \sigma_1^+ \sigma_1 - i \frac{\Gamma_2}{2} \sigma_2^+ \sigma_2
	\end{equation}
By reducing the master equation to a Schr\"{o}dinger equation with the non-Hermitian Hamiltonian $H_{\text{eff}}$, it becomes straightforward to solve for the system dynamics analytically. To this end, we denote the arbitrary single-excitation state vector as 
	\begin{equation}
\ket{\psi(t)} = c_{gg}(t) \ket{gg00} + c_{eg}(t) \ket{eg00} + c_{ge}(t) \ket{ge00} + c_{10}(t) \ket{gg10} + c_{01}(t) \ket{gg01}
	\end{equation}
where $\ket{i j n_1 n_2} \equiv \ket{i} \otimes \ket{j} \otimes \ket{n_1} \otimes \ket{n_2}$ represents the multipartite state in the order: qubit in node 1, qubit in node 2, cavity in node 1, cavity in node 2. The probability amplitudes evolve as
	\begin{equation}
	\begin{split}
i \dot{c}_{eg}(t) &= (\omega_{a1} - i \Gamma_1/2) c_{eg}(t) + g_1 c_{10}(t) \\
i \dot{c}_{ge}(t) &= (\omega_{a2} - i \Gamma_2/2) c_{ge}(t) + g_2 e^{-i\alpha} c_{01}(t) \\
i \dot{c}_{10}(t) &= [\omega_{c1} - i (\gamma_L+\gamma_R)/2] c_{10}(t) + g_1 c_{eg}(t) - i \gamma_l e^{ikD} c_{01}(t) \\
i \dot{c}_{01}(t) &= [\omega_{c2} - i (\gamma_L + \gamma_R)/2] c_{01}(t) + g_2 e^{i\alpha} c_{ge}(t) - i \gamma_R e^{ikD} c_{10}(t)
	\end{split}
	\end{equation}
To study entanglement generation, let us consider the initial state $\ket{\psi(0)} = \ket{eg00}$ such that $c_{eg}(0) = 1$ and all the other state coefficients are initially zero. Using this initial condition, the coefficients can be easily obtained by first taking the Laplace transform $f(t) \to \tilde{f}(p)$:
	\begin{equation}
	\begin{split}
ip \tilde{c}_{eg}(p) &= (\omega_{a1} - i \Gamma_1/2) \tilde{c}_{eg}(p) + g_1 \tilde{c}_{10}(p) + i \\
ip \tilde{c}_{ge}(p) &= (\omega_{a2} - i \Gamma_2/2) \tilde{c}_{ge}(p) + g_2 e^{-i\alpha} \tilde{c}_{01}(p) \\
ip \tilde{c}_{10}(p) &= [\omega_{c1} - i (\gamma_L+\gamma_R)/2] \tilde{c}_{10}(p) + g_1 \tilde{c}_{eg}(p) - i\gamma_L e^{ikD} \tilde{c_{01}}(p) \\
ip \tilde{c}_{01}(p) &= [\omega_{c2} - i (\gamma_L + \gamma_R)/2] \tilde{c}_{01}(p) + g_2 e^{i\alpha} \tilde{c}_{ge}(p) - i\gamma_R e^{ikD} \tilde{c}_{10}(p)
	\end{split}
	\end{equation}
To obtain an analytically tractable solution, we make a few simplifications to the setup: (1) assume perfect chirality such that $\gamma_L = 0$, (2) resonant condition such that $\omega_{c1} = \omega_{c2} = \omega_{a1} = \omega_{a2} = \omega_0$, (3) ignore atomic decay such that $\Gamma_1 = \Gamma_2 = 0$. With those simplifications, the probability amplitudes can be readily solved to give
	\begin{equation}
	\begin{split}
c_{eg} &= e^{-\frac{1}{4} t \left(\gamma _R+4 i \omega _0\right)} \left(\frac{\gamma _R \sinh \left(\frac{1}{4} \kappa_1 t \right)}{\kappa_1}+\cosh \left(\frac{1}{4} \kappa_1 t \right)\right) \\
c_{ge} &= \frac{4 g_1 g_2 \gamma _R \left(\kappa_1 \sinh \left(\frac{1}{4} \kappa_2 t \right)-\kappa_2 \sinh \left(\frac{1}{4} \kappa_1 t \right)\right) e^{\frac{1}{4} i \left(-4 \alpha +4 k D+i \gamma_{R} t -4 \omega _0 t\right)}}{\left(g_1^2-g_2^2\right) \kappa_1 \kappa_2} \\
c_{10} &= -\frac{4 i g_1 e^{-\frac{1}{4} t \left(\gamma _R+4 i \omega _0\right)} \sinh \left(\frac{1}{4} \kappa_1 t \right)}{\kappa_1} \\
c_{01} &= -\frac{i g_1 \gamma _R e^{i D k-\frac{1}{4} t \left(\gamma _R+4 i \omega _0\right)} \left(\gamma _R \left(\frac{\sinh \left(\frac{1}{4} \kappa_2 t \right)}{\kappa_2}-\frac{\sinh \left(\frac{1}{4} \kappa_1 t \right)}{\kappa_1}\right)+\cosh \left(\frac{1}{4} \kappa_1 t \right)-\cosh \left(\frac{1}{4} \kappa_2 t \right)\right)}{g_1^2-g_2^2}
	\end{split}
	\label{eq:coeffs}
	\end{equation}
where $\kappa_i = \sqrt{\gamma_R^2 - 16g_i^2}$. Note that although $c_{ge}$ and $c_{01}$ appears to be indeterminate when $g_1 = g_2$, the limits $\lim_{g_1 \to g_2} c_{ge}$ and $\lim_{g_1 \to g_2} c_{01}$ exist. Similarly, the limits when $\kappa_1 \to 0$, $\kappa_2 \to 0$ also exist. For example, in the case of $g_1 = g_2 = g$ (and hence $\kappa_1 = \kappa_2 = \kappa$), the corresponding probabilities are
	\begin{equation}
	\begin{split}
P_{eg} &= e^{-\frac{1}{2} \gamma _R t} \left|\frac{\gamma _R \sinh \left(\frac{1}{4} \kappa t \right)}{\kappa}+\cosh \left(\frac{1}{4} \kappa t\right)\right|^2\\
P_{ge} &= e^{-\frac{1}{2} \gamma _R t} \left|\frac{8 g^2 \gamma _R \left(\kappa t \cosh \left(\frac{1}{4} \kappa t \right)-4 \sinh \left(\frac{1}{4} \kappa t \right)\right) }{\kappa^3} \right|^2 \\
P_{10} &= e^{-\frac{1}{2} \gamma _R t} \left|\frac{4 g \sinh \left(\frac{1}{4} \kappa t \right)}{\kappa} \right|^2 \\
P_{01} &= e^{-\frac{1}{2} \gamma _R t} \left|\frac{2 g \gamma _R \left(\gamma _R \kappa t \cosh \left(\frac{1}{4} \kappa t \right) - \left(4\gamma_R + \kappa^2 t \right) \sinh \left(\frac{1}{4} \kappa t \right)\right)}{\kappa^3} \right|^2
	\end{split}
	\label{eq:probabilities}
	\end{equation}
Note that when $g \gtrsim \gamma_R / 2$, the probabilities begin to become oscillatory, consistent with the idea that Rabi oscillations occur when cavity-atom coupling rate exceeds dissipation rate. As a side note, by maximizing $P_{ge} (t)$ in Eq. (\ref{eq:probabilities}), we also obtain the optimal couplings for excitation transfer $\ket{eg00} \to \ket{ge00}$ to be $g_1 = g_2 \approx 0.43 \gamma_R$, which was numerically determined in Ref. \cite{mok2019long} for the entanglement transport of W-states. Since a single excited state can be mapped onto a W-state in an atomic ensemble, our calculations here can also be applied to cases with an atomic ensemble in each node, except that the optimal coupling should be modified by $g_{\text{opt}} = 0.43 / \sqrt{N}$, where $N$ is the number of atoms on each node. We have also checked that our analytical solution shows excellent agreement with the numerical results.

To quantify the generated entanglement, we calculate the concurrence between the qubits, after tracing out the cavity modes \cite{Zurek}. In our setup, the concurrence is given by the simple expression $C(t) = 2 |\rho_{12}(t)|$, where $\rho_{12}(t) = \ket{eg} \bra{ge} = c_{eg}(t) c_{ge}^* (t)$, with the coefficients obtained in Eq. (\ref{eq:coeffs}). It was demonstrated in Ref. \cite{gonzalez2015chiral} that by using a chiral waveguide to mediate the interactions between two qubits, one can obtain a maximum concurrence $C_{\text{max}} \equiv \max{C(t)} = 2/e \approx 0.735$. Here, we show that by adding cavities as an intermediary between the atoms and the chiral waveguide, the maximum concurrence can be significantly improved to give $C_{\text{max}} = 0.920$.
	\begin{figure}
\subfloat{%
  \includegraphics[width=0.333\linewidth]{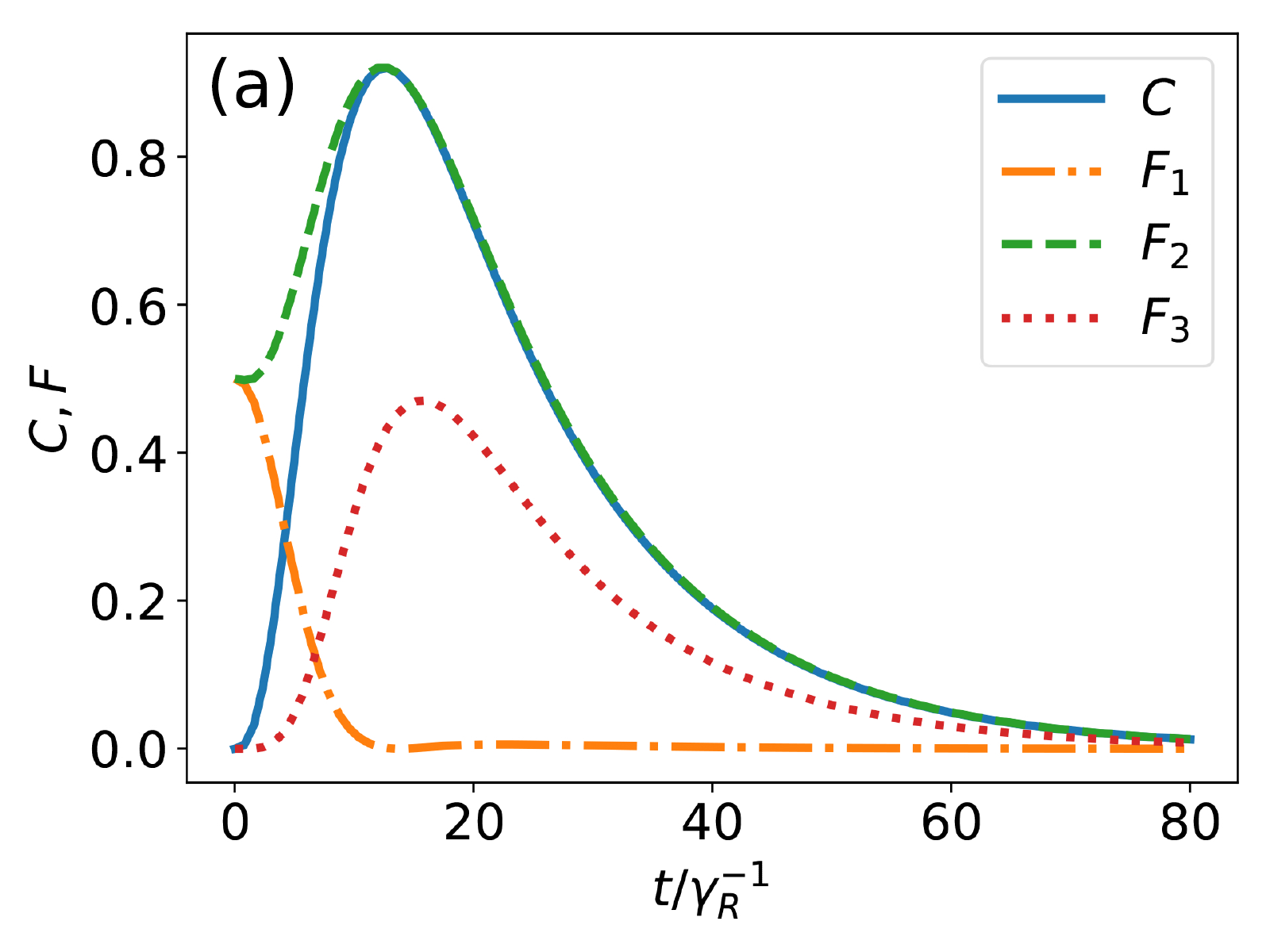}%
  \label{fig:conc_opt}%
}\hfill
\subfloat{%
  \includegraphics[width=0.333\linewidth]{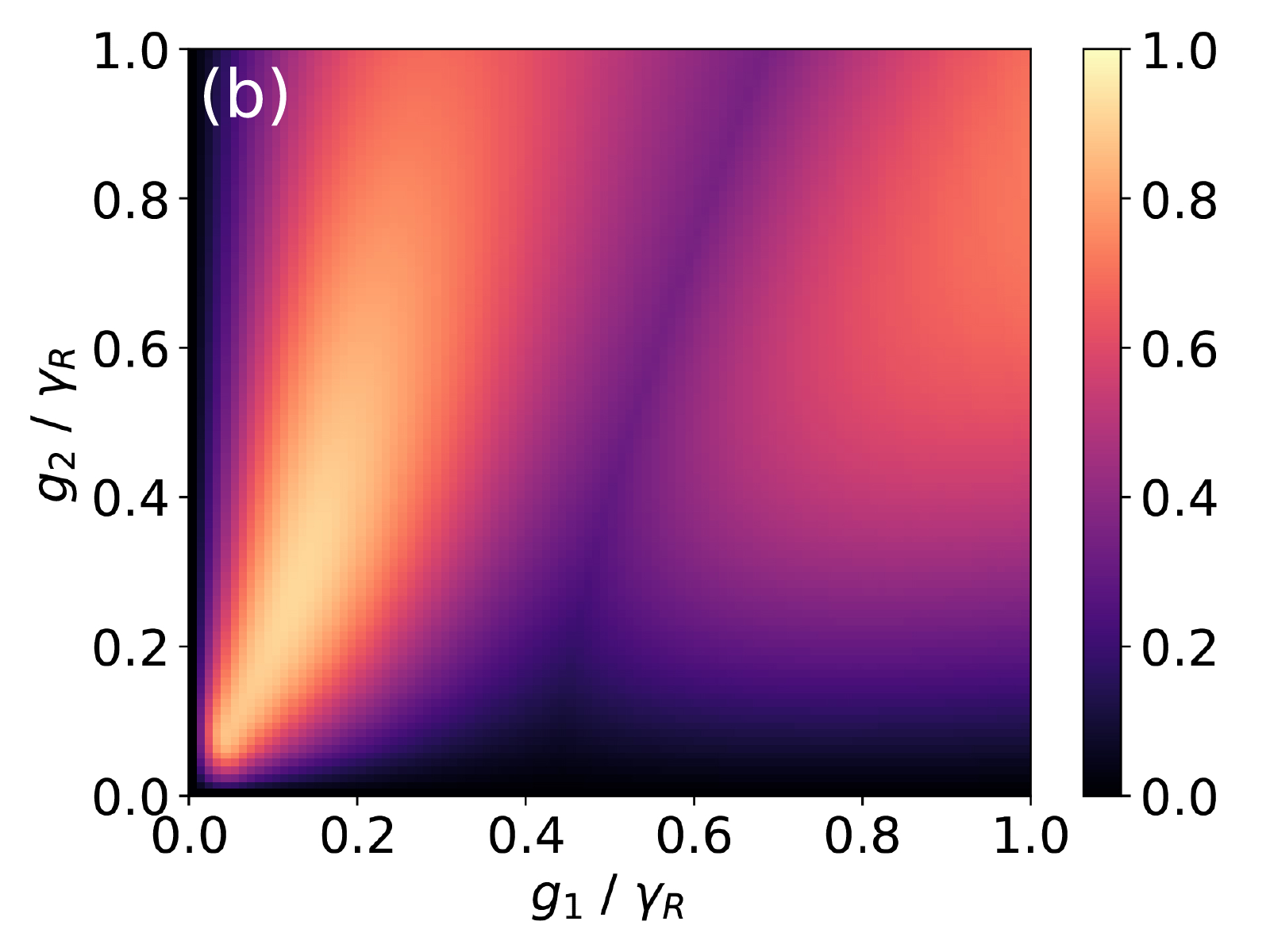}%
  \label{fig:cmax_g1g2}%
}\hfill
\subfloat{%
  \includegraphics[width=0.333\linewidth]{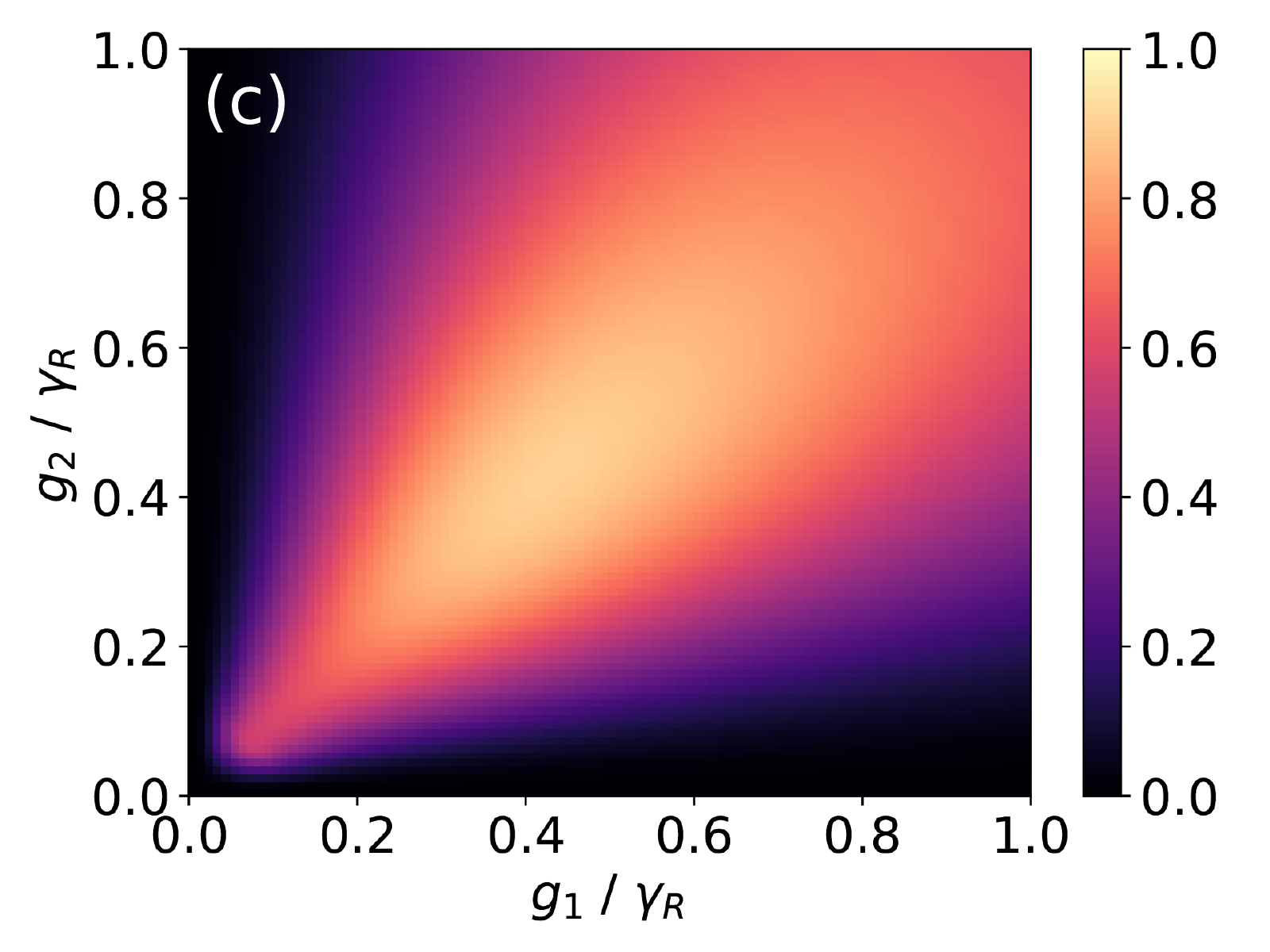}%
  \label{fig:transfermax_g1g2}%
}\hfill
	\caption{(a) Optimal entanglement generation for $g_1 \approx 0.126 \gamma_R$, $g_2 \approx 0.277 \gamma_R$. Various fidelities comparing the state with $\ket{\Psi^{\pm}}$ and $\ket{ge}$ are also plotted. (b) Maximum concurrence $C_{max}$ against $g_1$ and $g_2$. (c) Maximum transfer fidelity $\mathcal{F}_{3,\text{max}}$ against $g_1$ and $g_2$.}
	\label{fig:concurrence_comparison}
	\end{figure}
	
Fig. \subref*{fig:conc_opt} shows the concurrence $C(t)$ between the two qubits after optimizing over the system parameters. The optimal entanglement generation is achieved when $g_1 \approx 0.126 \gamma_R$ and $g_2 \approx 0.277 \gamma_R$, giving a maximum concurrence of $C_{\text{max}} = 0.920$. To better understand the dynamics, we also plot in Fig. \subref*{fig:conc_opt} the fidelities against different target states: $\mathcal{F}_i = \braket{\psi_i | \rho | \psi_i}$, with $\ket{\psi_1} = \ket{\Psi^+} = (\ket{eg} + \ket{ge})/\sqrt{2}$, $\ket{\psi_2} = \ket{\Psi^-} = (\ket{eg} - \ket{ge})/\sqrt{2}$ and $\ket{\psi_3} = \ket{eg}$. It is thus clear that the entangled state created is the odd Bell state $\ket{\Psi^-}$. One might also notice that the curves for $C$ and $\mathcal{F}_2$ are the same for $t > t_{\text{peak}}$, where $t_{\text{peak}}$ is the time where $C_{\text{max}}$ is attained. However, that only occurs at the optimal condition and is not true in general.

Plotting the maximum concurrence $C_{\text{max}}$ against $g_1$ and $g_2$ in Fig. \subref*{fig:cmax_g1g2}, it can be seen that good entanglement generation is achieved when $g_2 > g_1$. In the region near $g_1 = g_2 = 0.5 \gamma_R$, the entanglement between the two qubits is weak. This behaviour can be explained by the fact that when $g_1 \approx g_2$, the system performs a state transfer $\ket{eg} \to \ket{ge}$ instead of generating entanglement. Since a state transfer essentially maps a separable bipartite qubit state to another separable bipartite qubit state, it is thus expected that the concurrence will be low in such situations. Indeed, as Fig. \subref*{fig:transfermax_g1g2} shows, the maximum transfer fidelity $\mathcal{F}_{3,\text{max}}$ is the highest at $g_1 = g_2 = 0.43 \gamma_R$ as explained above. Intuitively, to generate the entangled state $\ket{\psi}$ with high concurrence, one has to engineer some kind of `partial state transfer', where only a fraction of the excitation in the first node (near 0.5) is transferred to the second node. This is also reflected in Fig. \subref*{fig:conc_opt} where $\mathcal{F}_{3}$ peaks at $0.42$ (less than 0.5 due to excitation leakage from the waveguide). Additionally, this also explains the asymmetrical couplings $g_1 \neq g_2$ since Fig. \subref*{fig:transfermax_g1g2} also shows that $g_1 = g_2$ is required for good excitation transfer. Thus, we conclude that the addition of cavities allows for greater control over the system dynamics via $g_1$ and $g_2$, resulting in a significant increase in concurrence from $0.73$ to $0.92$. 

\section{Role of imperfections}
\label{sec:imperfections}

Next, we investigate the effects of imperfections on our protocol, in order to better understand the experimental feasibility of our proposed scheme. From Fig. \subref*{fig:conc_opt}, we can already observe that $C_{\text{max}}$ is not highly sensitive to the values of $g_1$ and $g_2$, thus good entanglement generation can still be achieved even for imperfect couplings. Other than the couplings $g_1$ and $g_2$, we also study the effects of other imperfections such as: (1) distance between the nodes, (2) imperfect chirality, (3) detunings between the atoms and cavities and (4) spontaneous decay of atoms.

\subsection{Sensitivity to inter-nodal distance}

\begin{figure}
\subfloat{%
  \includegraphics[width=0.499\linewidth]{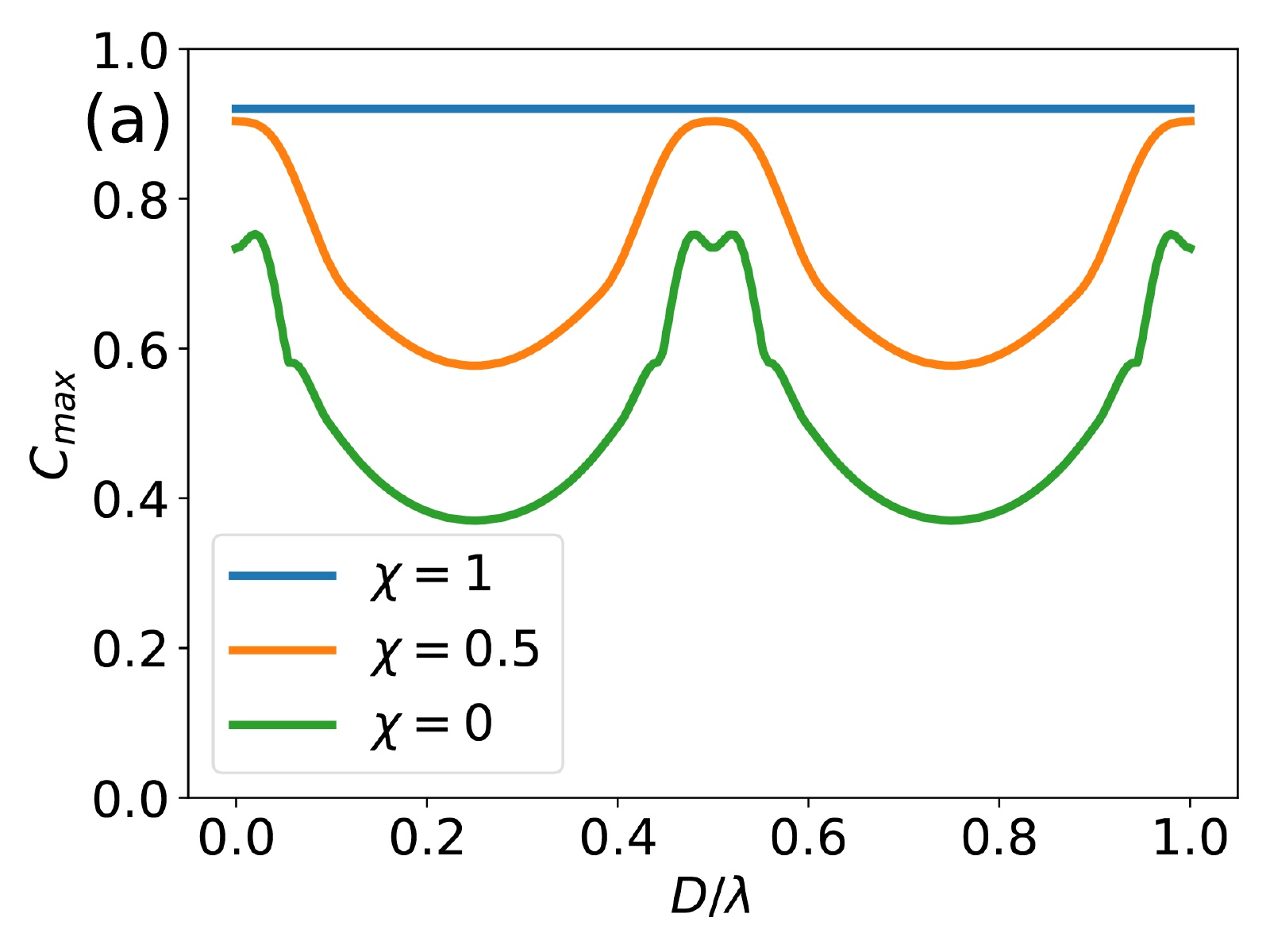}%
  \label{fig:cmax_d}%
}\hfill
\subfloat{%
  \includegraphics[width=0.499\linewidth]{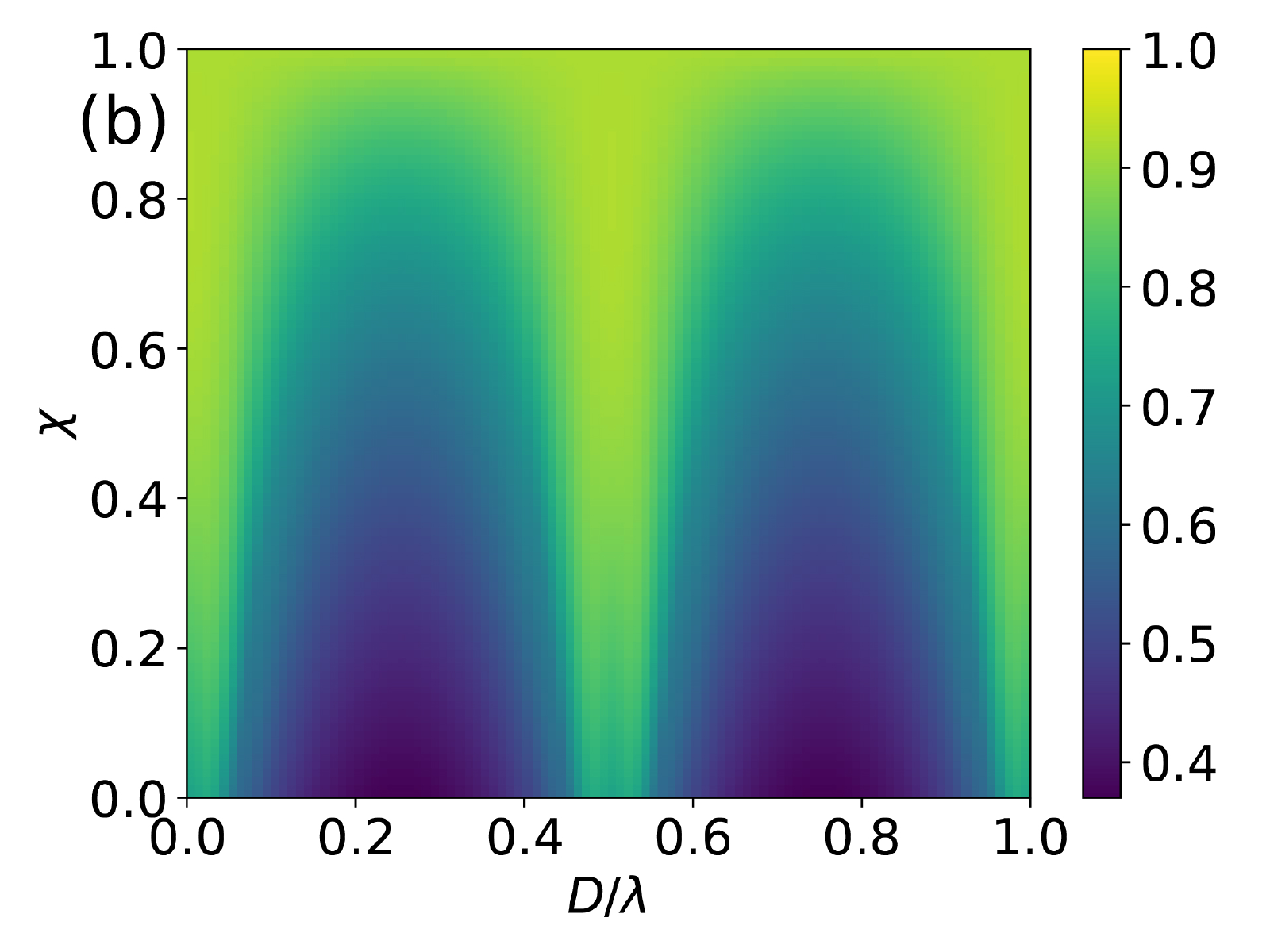}%
  \label{fig:cmax_d_chi}%
}\hfill
	\caption{(a) Variation of $C_{\text{max}}$ with inter-nodal distance $D$. (b) Variation of $C_{\text{max}}$ at $D = 0.5 \lambda$ with chirality $\chi$ (defined in main text). (c) $C_{\text{max}}$ against $D$ and $\chi$. In (a) and (b), the cavity-atom couplings are set at the optimal values $g_1 = 0.126 \gamma_R$, $g_2 = 0.277 \gamma_R$. }
	\label{fig:sensitivity_d}
	\end{figure}

Fig. \subref*{fig:cmax_d} shows the variation of $C_{\text{max}}$ with inter-nodal distance $D$, at the optimal condition shown in Fig. \subref*{fig:conc_opt}. Here we define chirality as $\chi = (\gamma_R - \gamma_L) / (\gamma_R + \gamma_L)$. In the ideal case of perfect chirality $\chi = 1$, $C_{\text{max}}$ is independent of the distance $D$ between the nodes. This is expected since having perfect chirality essentially realizes a cascaded system which is independent of the distance between the source and target subsystems ~\cite{PhysRevLett.70.2273,gardiner1993driving,carmichael2007statistical}. Without perfect chirality, the concurrence is maximised at $D = n \lambda / 2$, where $n$ is a non-negative integer and $\lambda$ is the wavelength of the photon in the waveguide. This resonant condition is essentially the same as that required for the localization of the photon wavepacket between the nodes as described in Ref. \cite{Gonzalez_Ballestero_2013}. Thus, an additional benefit of using a chiral waveguide is that the scheme becomes robust against fluctuations in the inter-nodal distance, on top of the improved entanglement.

\subsection{Imperfect chirality}

As mentioned, by having imperfect chirality, one has to place the nodes at the `sweet spot' $D = n \lambda / 2$ to achieve the best entanglement generation. For slightly imperfect chirality, $C_\text{max}$ remains relatively insensitive to distance fluctuations about the `sweet spot'. It is important to note that the Markovian assumption remains valid for small imperfections in chirality due to the dynamical nature of our scheme. More specifically, if the time delay between the nodes is much greater than the system evolution time, and assuming that the entanglement is utilised (or stored by decoupling the atoms from the cavity) at the peak timing $t_{\text{max}}$, then the $\gamma_L$ term simply contributes additional decay into the waveguide without introducing non-Markovian effects. Of course, this also means that the entanglement generation is bad for low chirality due to significant excitation leakage.

\subsection{Effects of detuning}

\begin{figure}
\subfloat{%
  \includegraphics[width=0.45\linewidth]{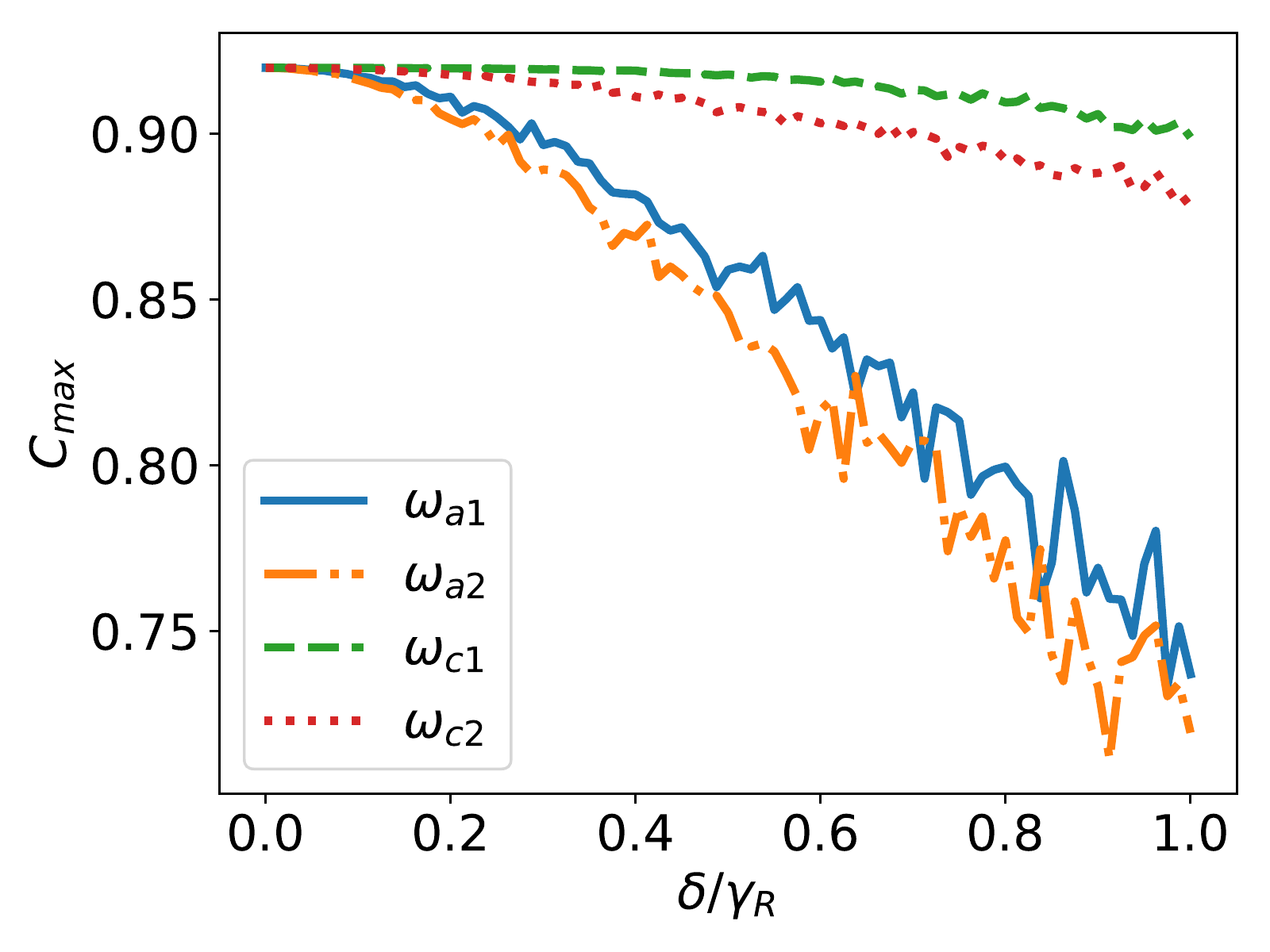}%
  \label{fig:cmax_fluctuation}%
}\hfill 
	\caption{Effect of random detunings on concurrence. The cavity-atom couplings are set at the optimal values $g_1 = 0.126 \gamma_R$, $g_2 = 0.277 \gamma_R$. }
	\label{fig:cmax_fluc}
	\end{figure}

Next, we consider the effects of detuning on entanglement generation. To do this, we first set $\omega_{a1} = \omega_{a2} = \omega_{c1} = \omega_{c2} = 1000 \gamma_R$, which is reasonable in the optical regime (for example, $\omega_{a1} \sim 10^{7} \gamma_R$ using the optical transition $F = 4 \to F^\prime = 5$ in cesium atoms \cite{PhysRevLett.102.083601}). To introduce detuning, we add a random fluctuation $\delta$ to the frequencies such that $\omega^\prime = \omega + \Delta \omega$ where $\omega^\prime$ is the new frequency and $\Delta \omega \in [-\delta/2, \delta/2]$ is a random variable with a uniform distribution. 
In Fig. \ref{fig:cmax_fluc}, we consider the average effect of random fluctuation on each of the four frequencies separately, using the same parameters as the optimal case shown in Fig. \subref*{fig:conc_opt}. Experimentally, a detuning of $\delta < \gamma_R$ can been achieved with microtoroidal resonators on photonic chips \cite{PhysRevLett.102.083601}. Thus, under realistic experimental conditions, the entanglement generation protocol is indeed robust against cavity and atom detunings. We also notice that fluctuations in cavity detuning has little effect on the concurrence compared to atom detuning. This is expected since the entanglement is stored in the atoms, thus having non-identical atoms will be more detrimental to the entanglement generation.

\subsection{Spontaneous decay of atoms}

In previous sections, we have neglected the spontaneous decay of atoms. Although the cavity linewidth $\gamma_R$ can be made much larger than the atomic linewidth, in practice there is still some decoherence effects due to atomic spontaneous emission into non-guided modes. Here, we consider the effects of non-zero atomic linewidth, which is assumed to be the same for both atoms, i.e. $\Gamma_1 = \Gamma_2 = \Gamma$. Fig. \ref{fig:cmax_decay} shows how the $C_{\text{max}}$ decreases with the atomic decay rate $\Gamma$. Note that if one uses the optimal conditions derived for $\Gamma = 0$, the maximum concurrence drops rapidly with increasing atomic decay, as shown by the $C_0$ curve in Fig. \ref{fig:cmax_decay}. However, one could first determine $\Gamma / \gamma_R$ of the setup to characterize the amount of loss, and calculate the optimal couplings for that particular setup. The resulting $C_{\text{max}}$ is depicted by the $C_{\text{opt}}$ curve in Fig. \ref{fig:cmax_decay}. With suitable optimization, the concurrence does not decrease as rapidly. For the example of cesium atoms in microtoroidal resonators (in the bad cavity regime), the atomic decay rate can be made to be $\Gamma < 0.1 \gamma_R$ \cite{PhysRevA.90.053822}, for which the entanglement is not greatly affected. However, compared to other imperfections, atomic decay is the most detrimental to entanglement generation. This is reasonable since the entanglement is stored in the excitation of the atoms, thus atomic losses have the most direct impact in destroying the entanglement between the atoms. 

	\begin{figure}
\subfloat{%
  \includegraphics[width=0.45\linewidth]{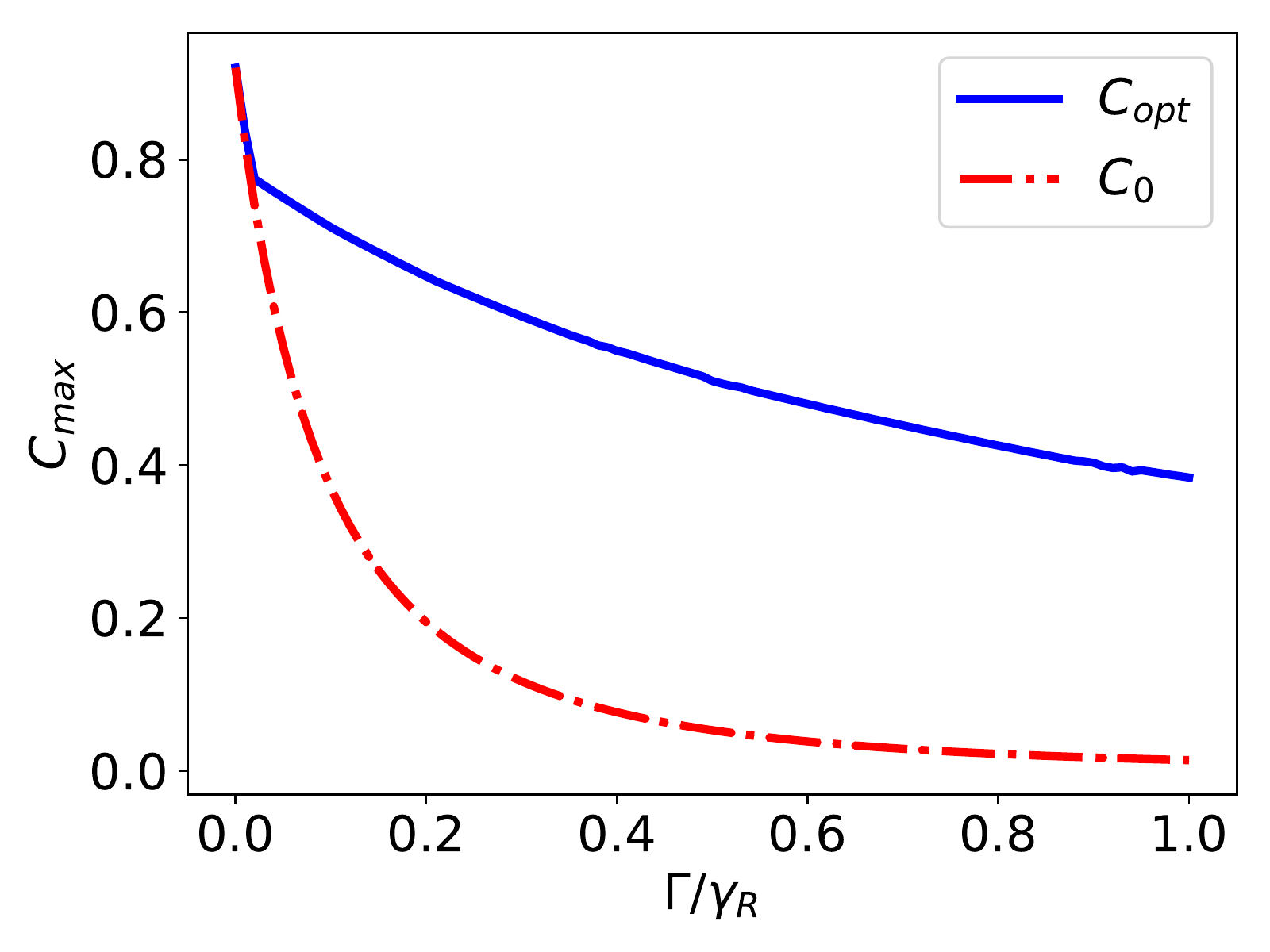}%
  \label{fig:cmax_decay}%
}\hfill
	\caption{Effect of atomic losses on concurrence. For the $C_0$ curve, the cavity-atom couplings are set at the optimal values $g_1 = 0.126 \gamma_R$, $g_2 = 0.277 \gamma_R$. For the $C_{\text{opt}}$ curve, separate optimization is done for different $\Gamma$.}
	\label{fig:cmax_decay}
	\end{figure}
	
\section{Chiral waveguide with $\gamma_{R1} \neq \gamma_{R2}$}
\label{sec:chiral_unequal}

So far, we have studied the chiral case where $\gamma_{R1} = \gamma_{R2} = \gamma_R$, in order to draw comparisons with the similar conditions set in Ref. \cite{gonzalez2015chiral}. If we relax this condition, the entanglement generation can be further enhanced. 

	\begin{figure}
\subfloat{%
  \includegraphics[width=0.499\linewidth]{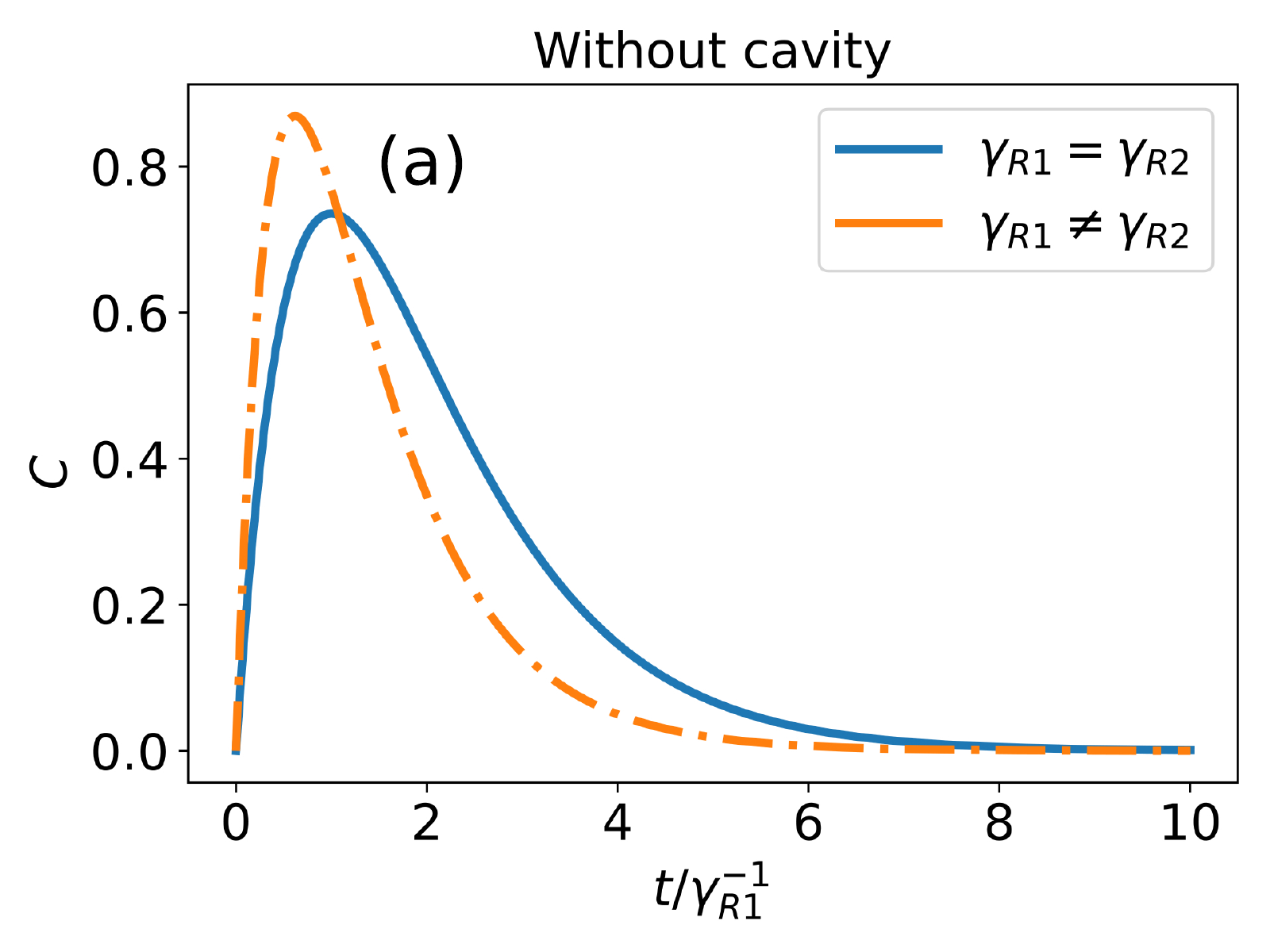}%
  \label{fig:cmax_unequal_nocavity}%
}\hfill
\subfloat{%
  \includegraphics[width=0.499\linewidth]{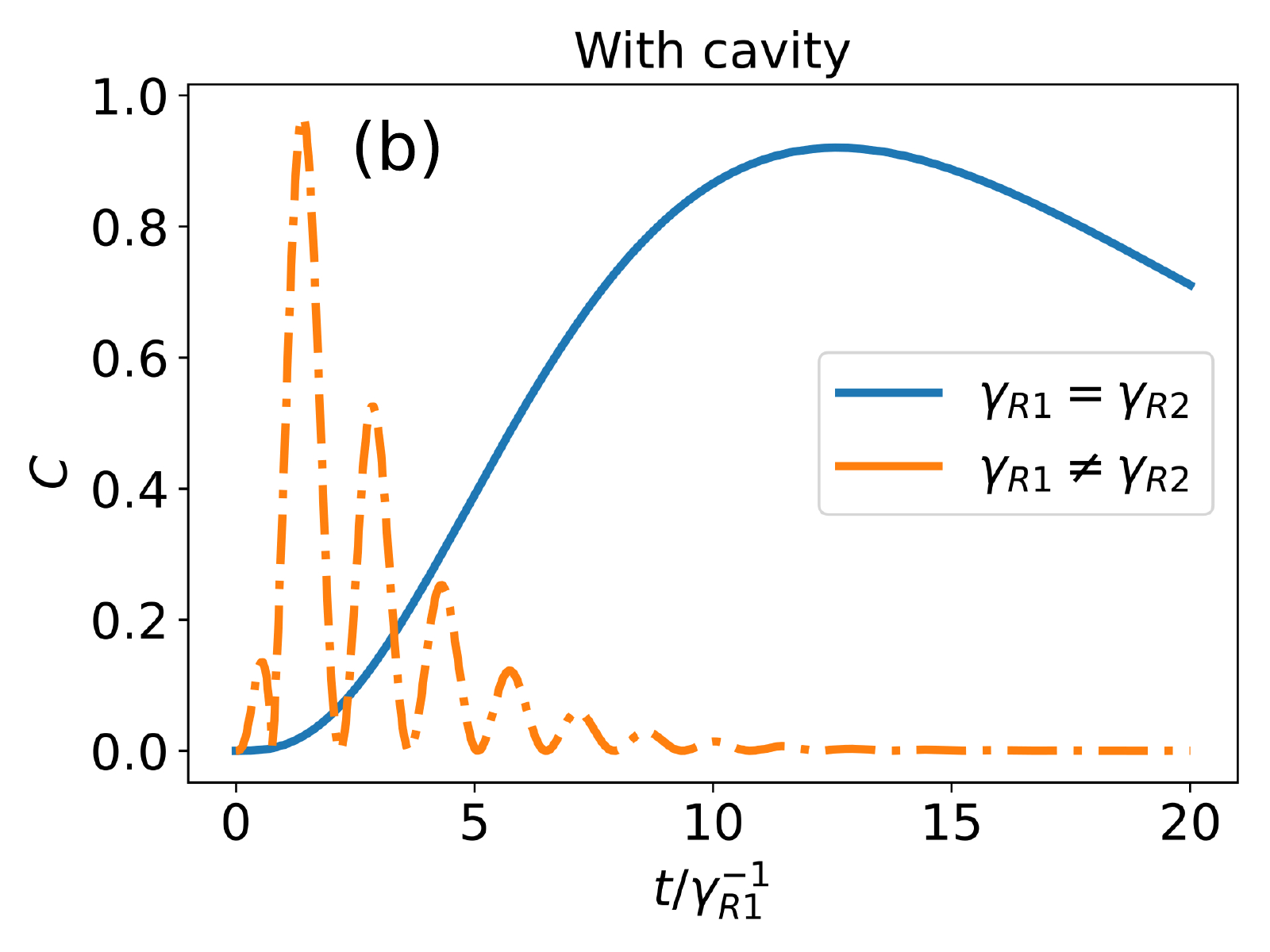}%
  \label{fig:cmax_unequal_withcavity}%
}\hfill
	\caption{Further enhancement in entanglement generation when $\gamma_{R1} \neq \gamma_{R2}$. (a) Optimal conditions for $\gamma_{R1} \neq \gamma_{R2}$ without cavity: $\gamma_{R2} = 3.88 \gamma_{R1}$. The solid blue curve is the result obtained in \cite{gonzalez2015chiral}. (b) Optimal conditions for $\gamma_{R1} \neq \gamma_{R2}$ with cavity: $\gamma_{R2} = 4.82 \gamma_{R1}$, $g_1 = 2.21 \gamma_{R1}$, $g_2 = 2.11 \gamma_{R1}$.}
	\label{fig:cmax_unequal}
	\end{figure}

Optimizing for the cases of $\gamma_{R1} = \gamma_{R2}$ and $\gamma_{R1} \neq \gamma_{R2}$ separately, we show in Fig. \ref{fig:cmax_unequal} that the optimized concurrence is indeed greater for $\gamma_{R1} \neq \gamma_{R2}$, both with and without cavities. For the case of no cavities, the optimal $C_{\text{max}}$ can be improved from $0.73$ (for $\gamma_{R1} = \gamma_{R2}$) as reported in Ref. \cite{gonzalez2015chiral} to $0.869$, obtained at $\gamma_{R2} = 3.88 \gamma_{R1}$. Similarly, for the case with cavities, we can further improve $C_{\text{max}}$ from the $0.92$ in \subref*{fig:conc_opt} to $0.968$, under the new optimal condition $g_1 = 2.21 \gamma_{R1}, g_2 = 2.11 \gamma_{R1}$ and $\gamma_{R2} = 4.82 \gamma_{R1}$. Noticeably, having $\gamma_{R1} \neq \gamma_{R2}$ in Fig. \subref*{fig:cmax_unequal_withcavity} causes more Rabi oscillations as a result of the larger optimal cavity-atom couplings $g_1$ and $g_2$. The intuition behind these asymmetrical couplings is similar to that explained earlier in Sec. \ref{sec:cavity_improve_entanglement}: good entanglement requires partial excitation transfer in order to establish the entangled Bell state. This requires asymmetrical couplings, for both inter-node couplings (characterised by $\gamma_{R1}$ and $\gamma_{R2}$) and intra-node couplings (characterised by $g_1$ and $g_2$ for the case with cavity). It should be noted however that tuning the cavity-waveguide interaction strengths $\gamma_{R1}$ and $\gamma_{R2}$ is more difficult in practice than tuning the cavity-atom couplings $g_1$ and $g_2$. Thus, even though the condition $\gamma_{R1} \neq \gamma_{R2}$ generates slightly better entanglement, simply setting $\gamma_{R1} = \gamma_{R2}$ with suitable cavity-atom couplings might be a more feasible scheme. Our results on entanglement generation in a chiral waveguide can thus be summarized in Tab. \ref{table:summary}. 

\begin{table}[htp]
\centering
\begin{adjustbox}{width=0.4\linewidth}
\begin{tabular}{|c|c|c|}
\hline
\multicolumn{1}{|l|}{} & \textbf{Without cavity} & \textbf{With cavity} \\ \hline
\textbf{$\gamma_{R1} = \gamma_{R2}$}             & 0.736                   & 0.920                \\ \hline
\textbf{$\gamma_{R1} \neq \gamma_{R2}$}             & 0.869                   & 0.969                \\ \hline
\end{tabular}
\end{adjustbox}
\caption{Optimal entanglement generation $C_{\text{max}}$ in chiral waveguides.}
\label{table:summary}
\end{table}

\section{Entanglement generation in non-chiral waveguide, $\gamma_L = \gamma_R$}
\label{sec:non_chiral}

\begin{figure}
\subfloat{%
  \includegraphics[width=0.333\linewidth]{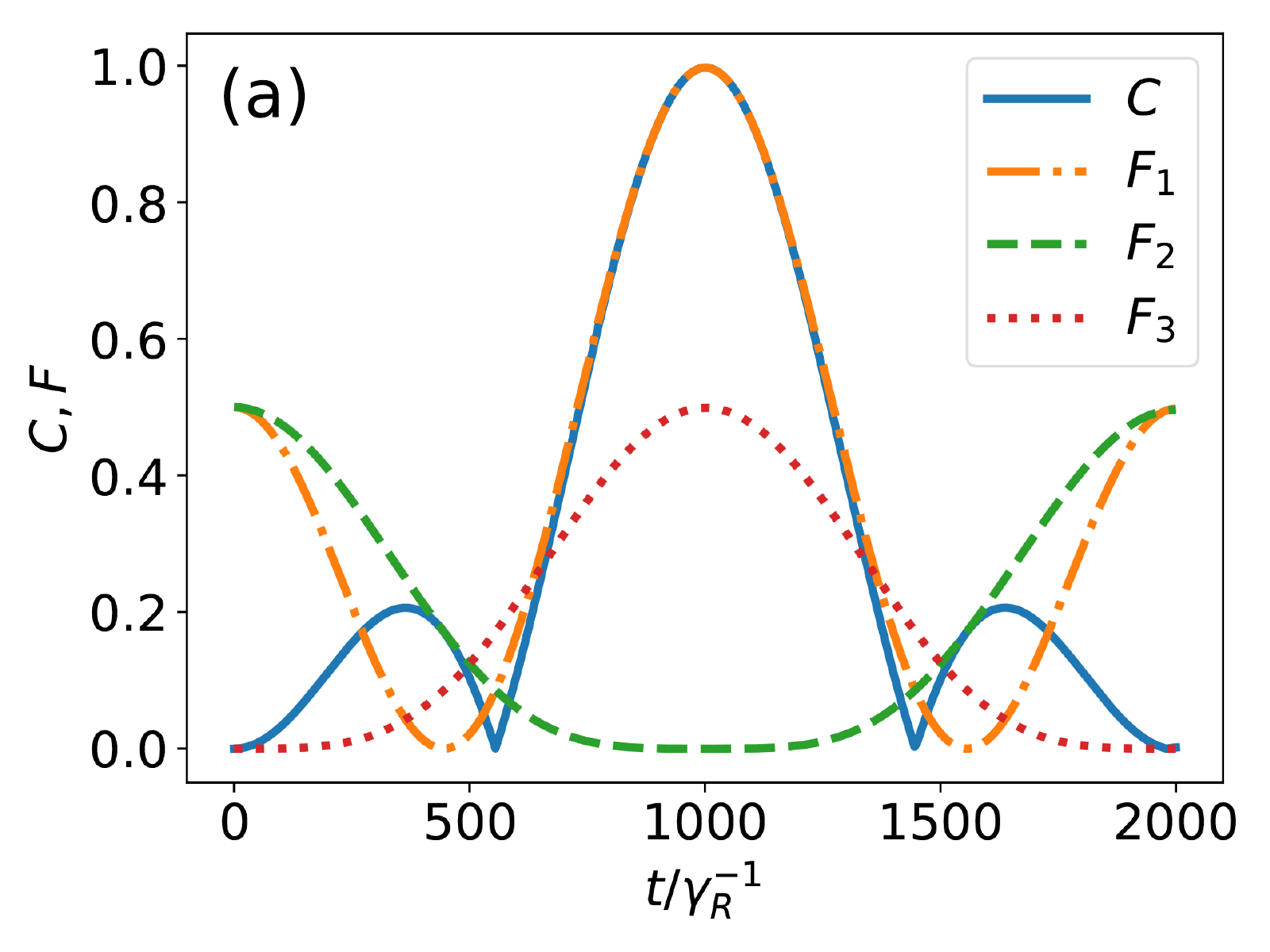}%
  \label{fig:conc_nonchiral}%
}\hfill
\subfloat{%
  \includegraphics[width=0.333\linewidth]{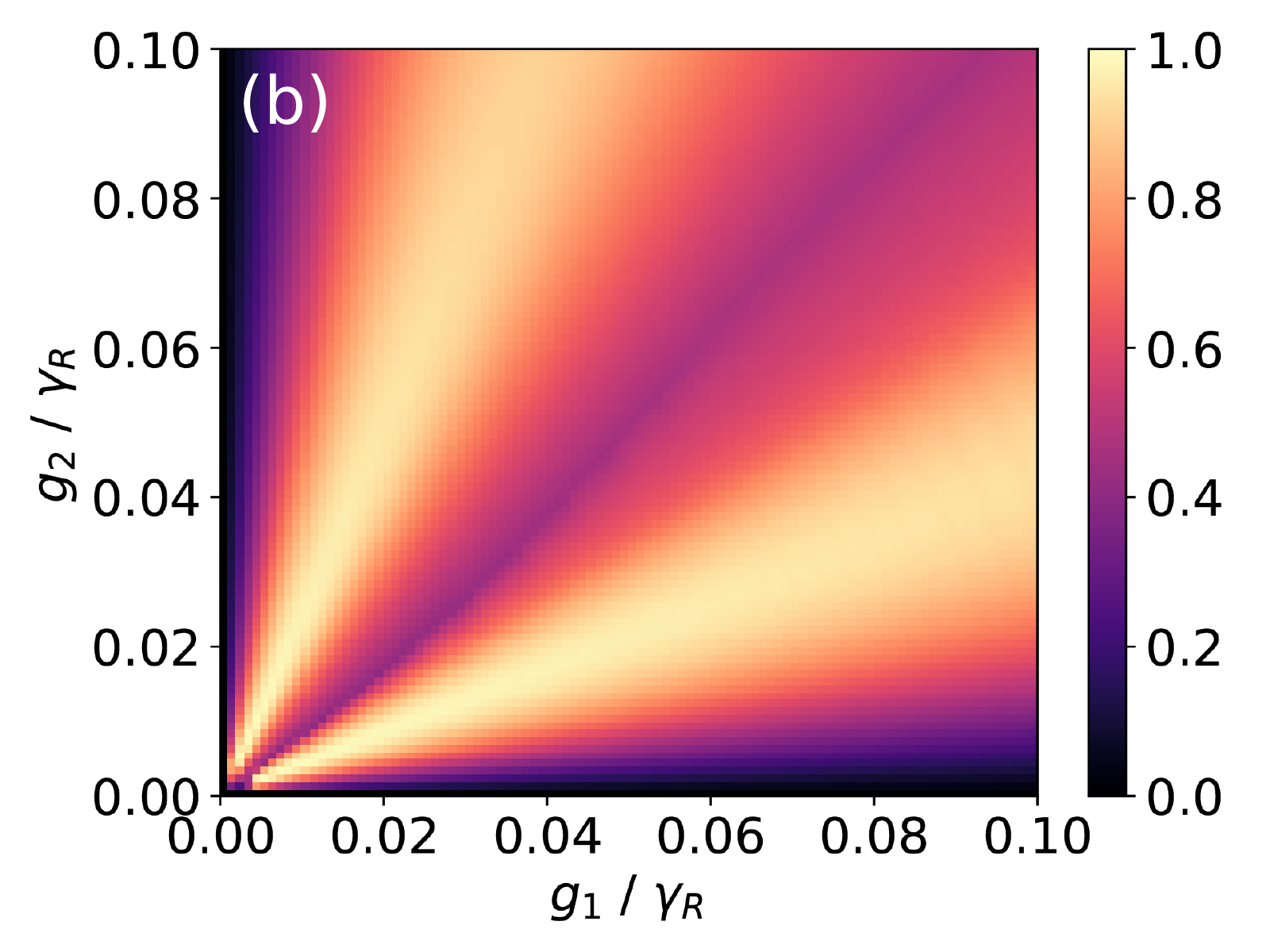}%
  \label{fig:cmax_g1g2_nonchiral}%
}\hfill
\subfloat{%
  \includegraphics[width=0.333\linewidth]{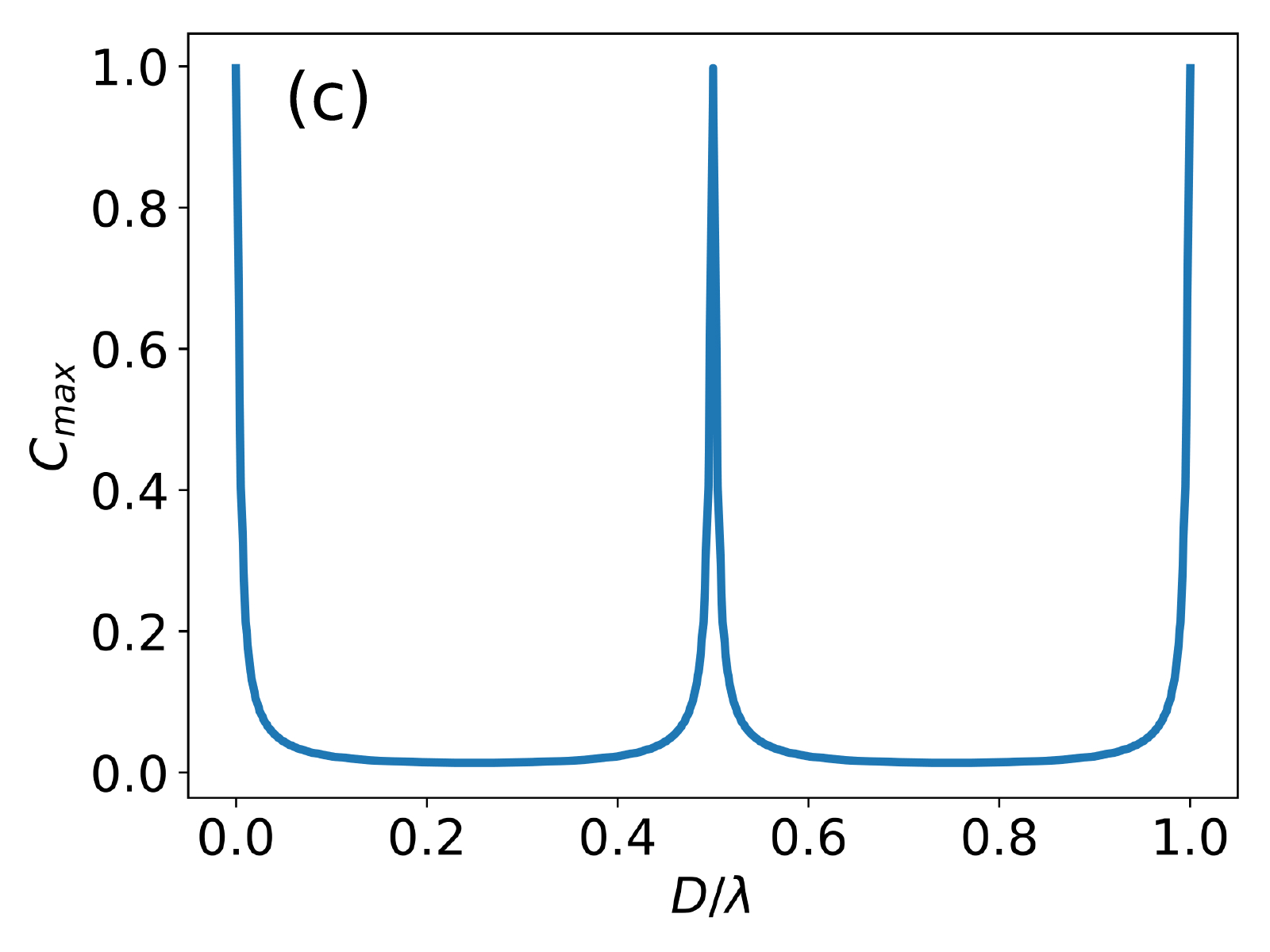}%
  \label{cmax_nonchiral_D}%
}\hfill
	\caption{ Non-chiral scheme for entanglement generation, with $\gamma_L = \gamma_R$. (a) Concurrence and various fidelities comparing the state with $\ket{\Psi^{\pm}}$ and $\ket{ge}$. (b) Maximum concurrence $C_{\text{max}}$ against $g_1$ and $g_2$. (c) $C_{\text{max}}$ against inter-nodal distance $D$. In (a) and (c), the optimal parameters $g_1 = 0.00410 \gamma_R, g_2 = 0.00170 \gamma_R$ are used.  }
	\label{fig:cmax_nonchiral}
	\end{figure}

To better understand the benefits of using a chiral waveguide, it is worthwhile to look at the optimal entanglement generation without chirality. In this section, we consider the case of a non-chiral waveguide where $\gamma_{L1} = \gamma_{R1} = \gamma_{L2} = \gamma_{R2} = \gamma$ and $\Gamma_i = 0$, $i=1,2$. Unsurprisingly, the optimal entanglement occurs again at the standing-wave condition $D = n\lambda / 2$, where the photon wavepacket is localised between the two nodes. This effect is stronger in the non-chiral case due to the greater symmetry with $\gamma_{R} = \gamma_{L}$. As such, the maximum concurrence can actually reach near unity with $C_{\text{max}} = 0.997$, shown in Fig. \subref*{fig:conc_nonchiral} with $D = \lambda/2$. Similar to Fig. \subref*{fig:conc_opt}, the $F_3$ curve peaks at $0.5$. Thus, the earlier intuition about the partial state transfer remains valid, and $g_1 \neq g_2$, as expected. The dependence of $C_{\text{max}}$ on $g_1$ and $g_2$ is depicted in Fig. \subref*{fig:cmax_g1g2_nonchiral}. Optimal entanglement is achieved with $g_1 = 0.00410 \gamma_R$ and $g_2 = 0.00170 \gamma_R$. Unlike the corresponding chiral case in Fig. \subref*{fig:cmax_g1g2}, here $C_{\text{max}}$ is invariant under the exchange $g_1 \leftrightarrow g_2$, due to the highly symmetrical setup. 

Note that while the chiral case generates the Bell state $(\ket{eg} - \ket{ge})/\sqrt{2}$ for any inter-nodal distance $D$, the non-chiral case actually generates $(\ket{eg} + (-1)^{n+1} \ket{ge})/\sqrt{2}$ at $D = n\lambda /2$ for non-negative integer $n$. This can be understood by examining the master equation. Setting $D = n \lambda / 2$ in Eq. (\ref{eq:lindblad_ME}), the coherent part of the waveguide-mediated interaction vanishes, giving (with $\gamma_R = \gamma_L = \gamma$)
	\begin{equation}
\dot{\rho} = -i [H_{\text{JC}}, \rho] + 2\gamma \mathcal{D} [a_1 + (-1)^{n} a_2] \rho
	\end{equation}
where the two cavities interact purely via dissipative coupling. The non-chiral scheme thus works on different physical mechanism than the chiral case, where both the coherent and incoherent couplings work together to generate entanglement. More importantly, there is a $\pi$-shift in the relative phase between $a_1$ and $a_2$ in the dissipator, which goes into the relative phase in the generated Bell state. Physically, the absence of chirality means that the phase $\exp(ikD)$ acquired from travelling between the nodes is non-negligible.

Although the non-chiral scheme can give near unity concurrence, there are several huge drawbacks with the non-chiral scheme. Firstly, the entanglement generation is much slower, as the optimal cavity-atom coupling rates are very small. Comparing with the chiral case, the peak timing is slower by 2 orders of magnitude. Taking into account realistic conditions for atomic decay rate, the entanglement will be negligible at the peak timing $t \approx 10^3 \gamma_R^{-1}$ as the atoms relaxes to the ground state $\ket{gg}$, which means that the non-chiral scheme is not feasible in practice. Moreover, as Fig. \subref*{cmax_nonchiral_D} shows, the entanglement is extremely sensitive to the inter-nodal distance as the scheme is highly reliant on the localisation of photon wavepacket which occurs at $D = n \lambda/2$. On top of that, the backflow of information from node 2 to node 1 also means that non-Markovian effects must be accounted into any long-distance implementation, which was shown in Ref. \cite{Gonzalez_Ballestero_2013} to worsen entanglement generation. These drawbacks can be aptly resolved by using a chiral waveguide, which can mediate fast dynamical entanglement generation with a low sensitivity to the inter-nodal distance.

\section{Conclusion}
\label{sec:conclusion}
%review of proposals NV entanglement
In this paper we propose a protocol for generating long-distance entanglement between the nodes of a chiral quantum network, where each node comprises a ring resonator coupled externally to an atom. Using an effective Hamiltonian approach we have obtained an analytical solution for the dynamics of the nodes, which shows a remarkable agreement with the numerical solution of the master equation. By optimising over the system parameters, we have demonstrated that the adding resonators to each node can improve the concurrence significantly from the $0.73$ reported in previous proposal \cite{gonzalez2015chiral} to $0.92$. This improvement is due to the additional level of control provided by the resonators over the system dynamics in order to achieve the partial excitation transfer essential for entanglement generation. Further improvement in concurrence of up to $C_{\text{max}} = 0.969$ can be obtained by relaxing the condition of $\gamma_{R1} = \gamma_{R2}$. Our proposal is also robust to experimental imperfections such as fluctuations in inter-nodal distance, cavity and atom detunings and spontaneous atomic decay. Particularly, we contrast our chiral protocol with the non-chiral case, where many of the severe drawbacks can be solved with a chiral waveguide.

The experimental feasibility of our protocol allows it to be easily implemented in state-of-the-art integrated photonic platforms, where photonic crystals and microtoroidal cavities can be integrated on the same chip in a scalable fashion. Moreover, our protocol has an added feature of operating in the weak coupling regime, allowing for easier implementation. Long distance light-matter entanglement has also been recently demonstrated experimentally in the optical regime over 50 km of optical fibre \cite{Krutyanskiy2019}. We also highlight that perfect chirality is not required for our scheme to work well. We believe that our findings will give further impetus to the practical realization of quantum networks.

In addition, we would like to point out that there have been several theoretical proposals on coupling diamond NV centres with ring cavities for generating entangled states between the colour centres \cite{Shi2010,Yang1,Yang2,Liu2013}. Since colour centres are solid-state systems, there is no need to trap them as it is the case with cold atoms. It might thus be interesting to implement our proposal by coupling ring resonators with colour centers which are connected together via the chiral quantum channel.

\bibliographystyle{apsrev4-1}
\bibliography{Biblio_v1}
\onecolumngrid

\appendix

\end{document}